\documentclass{JCIN22}
\usepackage{graphicx}
\usepackage[numbers]{gbt7714}
\begin{document}

\ArticleType{Review paper}
\Year{2022}

\title{6G-enabled Edge AI for Metaverse: Challenges, Methods, and Future Research Directions}

\author[1]{Luyi Chang}
\author[1]{Zhe Zhang}
\author[1]{Pei Li}
\author[1]{Shan Xi}
\author[1]{Wei Guo}
\author[2]{Yukang Shen}
\author[3]{Zehui Xiong}
\author[4]{Jiawen Kang}
\author[5]{Dusit Niyato}
\author[6]{Xiuquan Qiao}
\author[1,*]{Yi Wu}
\maketitle

{\bf\textit{Abstract---} 6G-enabled edge intelligence opens up a new era of Internet of Everything and makes it possible to interconnect people-devices-cloud anytime, anywhere. More and more next-generation wireless network smart service applications are changing our way of life and improving our quality of life. {As the hottest new form of next-generation Internet applications, Metaverse is striving to connect billions of users and create a shared world where virtual and reality merge.} However, limited by resources, computing power, and sensory devices, Metaverse is still far from realizing its full vision of immersion, materialization, and interoperability. To this end, this survey aims to realize this vision through the organic integration of 6G-enabled edge AI and Metaverse. Specifically, we first introduce three new types of edge-Metaverse architectures that use 6G-enabled edge AI to solve resource and computing constraints in Metaverse. Then we summarize technical challenges that these architectures face in Metaverse and the existing solutions. Furthermore, we explore how the edge-Metaverse architecture technology helps Metaverse to interact and share digital data. Finally, we discuss future research directions to realize the true vision of Metaverse with 6G-enabled edge AI.\\[-1.5mm]

}

\barefootnote{\textsuperscript{1}Luyi Chang, Zhe Zhang, Pei Li, Shan Xi, Wei Guo, and Yi Wu are with School of Data Science and Technology, Heilongjiang University, Harbin 150080, China (e-mail: 2202518@s.hlju.edu.cn; 2202517@s.hlju.edu.cn; 2212623@s.hlju.edu.cn; 2191831@s.hlju.edu.cn; 20203548@s.hlju.edu.cn; 1995050@hlju.edu.cn). \textsuperscript{2}Yukang Shen is with SenseTime Group Limited, Shenzhen 518000, China (e-mail: shenyukang1@sensetime.com). \textsuperscript{3}Zehui Xiong is with Pillar of Information Systems Technology and Design, Singapore University of Technology and Design, Singapore 487372, Singapore (e-mail: zehui\_xiong@sutd.edu.sg). \textsuperscript{4}Jiawen Kang is with the School of Automation, Guangdong University of Technology, China (e-mail: kavinkang@gdut.edu.cn). \textsuperscript{5}Dusit Niyato is with the School of Computer Science and Engineering, Nanyang Technological University, Singapore (e-mail: dniyato@ntu.edu.sg). \textsuperscript{6}Xiuquan Qiao is with State Key Laboratory of Networking and Switching Technology, Beijing University of Posts and Telecommunications, Beijing 100876, China (e-mail: qiaoxq@bupt.edu.cn).\\\indent
\textsuperscript{*}Yi Wu is the corresponding author.}

\section{INTRODUCTION}
Data is considered the new oil, as it is the fuel powering artificial intelligence (AI)\textsuperscript{\cite{9663104}} and creates tremendous value for many domains from business activities to scientific research\textsuperscript{\cite{jia2019towards}}. Data-driven AI technology has spawned a series of emerging technologies such as computer vision\textsuperscript{\cite{szeliski2010computer}}, natural language processing\textsuperscript{\cite{chowdhary2020natural}}, and data mining\textsuperscript{\cite{han2011data}}. These emerging technologies have revolutionized AI technology and greatly improved people's quality of life. However, with the rapid development of information technology and mobile device technology, the need to deploy AI technology on devices is becoming more and more urgent\textsuperscript{\cite{liu2020federated}}. In this context, industry, and academia have developed a new learning paradigm, edge artificial intelligence (Edge AI)\textsuperscript{\cite{letaief2021edge}}, which allows AI models to be deployed on devices and perform real-time data processing and model inference. The emergence of Edge AI further expands the application scope of AI and gradually opens a new era of ``Internet of Everything''. Nevertheless, Edge AI paradigm and its applications still have the following issues that need to be optimized:
\begin{itemize}
    \item \textbf{High Latency:} Since edge AI generally involves thousands of remote devices and needs to transmit and process massive amounts of data\textsuperscript{\cite{wang2019edge,lim2020federated}}, the high latency issue in the current network environment has always been one of the bottlenecks hindering the wide application of edge AI\textsuperscript{\cite{wang2020convergence,liu2021resource}}. 
    {In particular, some haptic AI applications require a transfer rate of at least $1$ $Mbit/s$ and a latency of no more than $1$ $ms$\textsuperscript{\cite{ivkovic2015quantifying}}.}
    Although the fifth generation (5G) of mobile communication has alleviated this problem to a great extent, the exponential growth of the volume of data and models forces us to require a faster and more stable new generation of mobile communication technology\textsuperscript{\cite{liu2020toward}}.
    \item \textbf{Fragile Stability:} In edge AI, the training of large-scale models often requires powerful computing power and stable network connections, especially the training of large language models\textsuperscript{\cite{mikolov2011strategies}}. However, the current network environment is only suitable for the training of small-scale models\textsuperscript{\cite{li2021terapipe}}. This is because the fragility of the network connection leads to the failure of large-scale model training.
    \item \textbf{Low Security:} The current network architecture no longer meets the security needs of thousands of remote devices connecting to cloud servers today\textsuperscript{\cite{wang2019edge}}. Furthermore, the openness of the network further challenges the security of the current network architecture.
\end{itemize}

To address the above problems, the 6G-enabled edge AI paradigm came into being, which is the product of the efficient integration and collaboration of AI technology and 6G mobile communication technology. 6G mobile communication technology provides edge AI with lower latency, more stable network connection, and a more secure network architecture. Specifically, the 6G wireless network uses the terahertz (THz) frequency band with a peak rate of $1T b/s$ to achieve a network delay with a transmission rate of less than $1ms$\textsuperscript{\cite{saad2019vision}}. Open radio access network (RAN) is an emerging framework for network transfer through infrastructure virtualization and embedded intelligence to provide end users with more stable network connectivity services and advanced capabilities\textsuperscript{\cite{letaief2019roadmap}}. In addition, the advanced network slicing architecture in 6G technology provides users with more intelligent and secure network services~\cite{liu2020federated}. Therefore, the 6G-enabled edge AI architecture leads the development of next-generation wireless network intelligent applications.

Undoubtedly, the Metaverse is one of the most concerned and promising smart applications in the next generation of wireless network intelligent applications~\cite{lee2021all,wang2022survey,gadekallu2022blockchain}. {The goal of the Metaverse is to create a world where virtuality and reality merge that can accommodate hundreds of millions of people interacting online\textsuperscript{\cite{duan2021metaverse}}.} This means that the Metaverse places higher demands on the current edge AI architecture. From the perspective of network or communication, the key requirements of the Metaverse for edge AI are bandwidth, latency, reliability, and intelligence. That is to say, the Metaverse needs mobile communication platforms to provide ultra-high bandwidth, ultra-low latency, ultra-high reliability, and more intelligent services\textsuperscript{\cite{xu2022full}}. Different from traditional intelligent applications, Metaverse infrastructure involves network architecture, a communication platform, virtual technology, hardware facilities, intelligent algorithms, etc. In addition, the Metaverse involves human senses, psychology, thought, and morals. From a high level view, the Metaverse is a very complex virtual world. Therefore, we need to explore how the 6G-enabled edge AI architecture can serve and empower the Metaverse.

In this paper, we survey the challenges, methods, and future research directions of the 6G-enabled edge AI architecture empowered by Metaverse. Previous surveys are shown in Tab. \ref{survey}, such as~\cite{wang2022survey,lee2021all,ning2021survey,pham2022artificial}, focused on Metaverse challenges, technologies, applications, technological singularities, virtual ecosystems, and research agendas. More specifically, these surveys focus more on Metaverse-related technologies rather than other technologies enabling them. In fact, the technical architecture related to the Metaverse is the focus and difficulty of current research. Therefore, we are now turning to edge AI architectures related to the Metaverse and seeking efficient integration and collaboration between the two. The closest to our survey is reference\textsuperscript{\cite{xu2022full}}, but they focus on the decentralized Metaverse architecture based on blockchain technology and its economic system. Instead, our survey focuses on Edge Cloud-Metaverse architecture, Mobile Edge Cloud-Metaverse architecture, and Decentralized-Metaverse architecture. In addition, we also summarize the challenges and future research directions of 6G-enabled edge AI in the Metaverse. The contributions of our survey are as follows:
\begin{itemize}
    \item We first give the definitions of 6G, edge intelligence, and Metaverse and give an analysis of the fusion of 6G-enabled edge AI and Metaverse. In particular, we also provide a tutorial that focuses on the features, architecture, technology, and applications of the Metaverse. With this concise tutorial, readers are able to stay abreast of cutting-edge developments in this topic.
    \item We discuss the types of architectures in the 6G-enabled edge AI empowered Metaverse and investigate solutions to implement the above architectures. Unlike traditional edge AI architectures, the edge AI-based Metaverse architectures involve data interaction and transformation between the virtual world and the real world. We then summarize the key challenges and existing methods for the fusion of such technical solutions with the Metaverse. Through this, readers can understand how the future development of 6G-enabled edge AI will play an important role in empowering Metaverse.
    \item We outline future research directions to pave the way for research attempts to empower Metaverse with 6G-enabled edge AI. Specifically, this survey aims to help researchers gain an in-depth understanding of the challenges, methods, and future research questions that edge AI empowers the Metaverse.
\end{itemize}

The remainder of this paper is organized as follows. Sec. \ref{sec-2} describes the concepts and features of 6G and 6G-enabled edge intelligence. Sec. \ref{sec-3} introduces the Metaverse, including features, architecture, applications, and history. Sec. \ref{sec-4} discuss the types of architectures in the 6G-enabled edge AI-empowered Metaverse, then summarizes the key challenges and existing methods. Future research questions are presented in Sec. \ref{sec-5} and conclusions are drawn in Sec. \ref{sec-6}.

\begin{table*}[!t]
\scriptsize
\centering
\caption{\textcolor{black}{Summary of related work and our survey.}}\label{survey}
\begin{tabular}{|c|l|l|}
\hline
\textbf{Reference} &
  \multicolumn{1}{c|}{\textbf{Key focus of survey}} &
  \multicolumn{1}{c|}{\textbf{How our survey differ}} \\ \hline
\cite{lee2021all} &
  \begin{tabular}[c]{@{}l@{}}Discuss threats to the security and privacy of the Metaverse and \\ present key challenges for the Metaverse system.\end{tabular} &
  \begin{tabular}[c]{@{}l@{}}Our survey discusses the types of architectures in the 6G-enabled edge AI empowered\\ Metaverse and investigate solutions to implement the above architectures.\end{tabular} \\ \hline
\cite{wang2022survey} &
  \begin{tabular}[c]{@{}l@{}}Discuss six user-centric factors in the Metaverse ecosystem: avatar,\\ content creation, virtual economy, social acceptability, security and\\ privacy, and trust and responsibility.\end{tabular} &
  \begin{tabular}[c]{@{}l@{}}Instead of focusing on the ecosystem, we focus on how the technical architecture of\\ the Metaverse is implemented in the edge network.\end{tabular} \\ \hline
\cite{ning2021survey} &
  \begin{tabular}[c]{@{}l@{}}Introduce the development status of Metaverse from five aspects: network\\ infrastructure, management technology, basic general technology, virtual\\ reality object connection and virtual reality fusion.\end{tabular} &
  \multirow{2}{*}{\begin{tabular}[c]{@{}l@{}}Our survey analyzes the fusion of 6G-enabled edge AI and Metaverse, and introduces\\ the features, architecture, technologies, and applications of Metaverse.\end{tabular}} \\ \cline{1-2}
\cite{pham2022artificial} &
  \begin{tabular}[c]{@{}l@{}}Discuss the role of artificial intelligence in the infrastructure and \\ development of the Metaverse.\end{tabular} & \\ \hline
\cite{xu2022full} &
  \begin{tabular}[c]{@{}l@{}}The focus is on the decentralized Metaverse architecture based on \\ blockchain technology and its economic system.\end{tabular} &
  \begin{tabular}[c]{@{}l@{}}Our survey focuses on Edge Cloud-Metaverse architecture, Mobile Edge Cloud-Metaverse\\ architecture, and Decentralized-Metaverse architecture.\end{tabular} \\ \hline
\end{tabular}
\end{table*}

\section{6G \& EDGE INTELLIGENCE}\label{sec-2}
\subsection{6G}
6-th generation (6G) mobile communication, as a continuation and enhancement above 4/5G mobile communication technology, is an advanced communication technology that integrates sensing, storage, communication, control, and computing capability. 
Compared with 4/5G, 6G has many advantages such as high performance, global coverage, real-time processing, high reliability, and energy efficiency\textsuperscript{\cite{saad2019vision}}.

Each time the mobile communication technology is updated and iterated, its various performance indicators are improved by $10$ to $100$ times compared to the previous generation.
As shown in Tab. \ref{tab-2}, compared with 5G, 6G has more than $100$ times improvement in peak transmission rate, reliability, traffic density, positioning accuracy, and connection density.
For example, 6G wireless networks use a terahertz band with a $1T b/s$ peak rate and less than $1$ $ms$ network latency in terms of transmission rate, which significantly improves the quality of experience (QoE) for users. 
6G can accurately locate within an activity range of $1$ $m$ outdoors and $10$ $cm$ indoors in positioning accuracy.
Moreover, in terms of coverage, the convergence of traditional terrestrial communication facilities, satellites, and unmanned aerial vehicles (UAVs) will enable 6G networks to provide global coverage, whether land or sea.
More importantly, in the field of security, compared with the traditional plug-in and patch-type network security mechanisms, 6G adopts endogenous security technology to have a stronger ability to resist unknown security threats.
It can be seen that the updated performance indicators of 6G provide good technical support for technology industries such as computer vision, blockchain, AI, the Internet of Things, robotics, and user interaction. In addition, driven by AI technology, 6G will provide the following three new service types\textsuperscript{\cite{9205981}}.

\begin{table}[!t]
\scriptsize
	\centering
	\caption{\textcolor{black}{Performance indicators comparison between 6G and 5G.}}
	\begin{tabular}{|c|c|c|}\hline
		\textbf{Performance Indicators} & \textbf{5G} & \textbf{6G}\\\hline
		Peak transmission rate & 10 $\sim$ 20 $Gb/s$ & 100 $Gb/s$ $\sim$ 1 $Tb/s$\\
		User experience rate & 0.1 $\sim$ 1 $Gb/s$ & 30 $\sim$ 50  $Gb/s$\\
		Time delay & 10 $\sim$ 50 $ms$ & 0.1 $\sim$ 1 $ms$\\
		Reliability & $10^{-5}$ & $10^{-9}$\\
		Flow density & 10 $Tbps/km^{2}$ & 100 $\sim$ 1000 $Tbps/km^{2}$\\
		Positioning precision & 1 $\sim$ 10 $m$ & 0.1 $\sim$ 1 $m$\\
		Connection density & 1 $million/km^{3}$ & 10 $\sim$ 100 $million/km^{3}$\\
		Network efficiency & 100 $bit/s$ & 200 $bit/s$\\
		Mobility & 500 $km/h$ & 1000 $km/h$\\
		Spectrum bandwidth & 30 $\sim$ 100 $bps/Hz$ & 200 $\sim$ 500 $bps/Hz$\\	
		Base station computing power & 100 $\sim$ 200 $Tops$ & 1000  $Tops$\\
		Coverage & Partial & Global\\
		Security & Patchy security & Endogenous security\\
		Information timeliness & High & Extremely high\\\hline
	\end{tabular}
	\label{tab-2}
\end{table}


\textit{1) New Media:} With the rapid development of virtual reality technology, the form of information interaction is constantly evolving, which accelerates the large-scale application of multi-sensory extended reality (XR) and holographic imaging technology. In virtual interaction, 6G can continuously provide users with a near-zero-latency sensory interconnection experience, such as the user's virtual movement in the Metaverse, virtual meetings, virtual paintings, wireless brain-computer interaction, and other interactive, immersive holographic experiences.

\textit{2) New Services:} 6G is the main driver of multiple new service models. For example, 6G communication technology provides users with precise service models in autonomous driving, industrial control, telemedicine, Internet robots, and autonomous systems, bringing a more convenient lifestyle.

\textit{3) New Infrastructure:} 6G infrastructure mainly includes information infrastructure, fusion infrastructure, and innovation infrastructure. In particular, the 6G communication system integrates infrastructure such as ground, UAV, and satellite Internet and features high bandwidth, low latency, strong reliability, and global coverage.

\begin{figure*}[!t]
	\centering
	\includegraphics[width=0.7\linewidth]{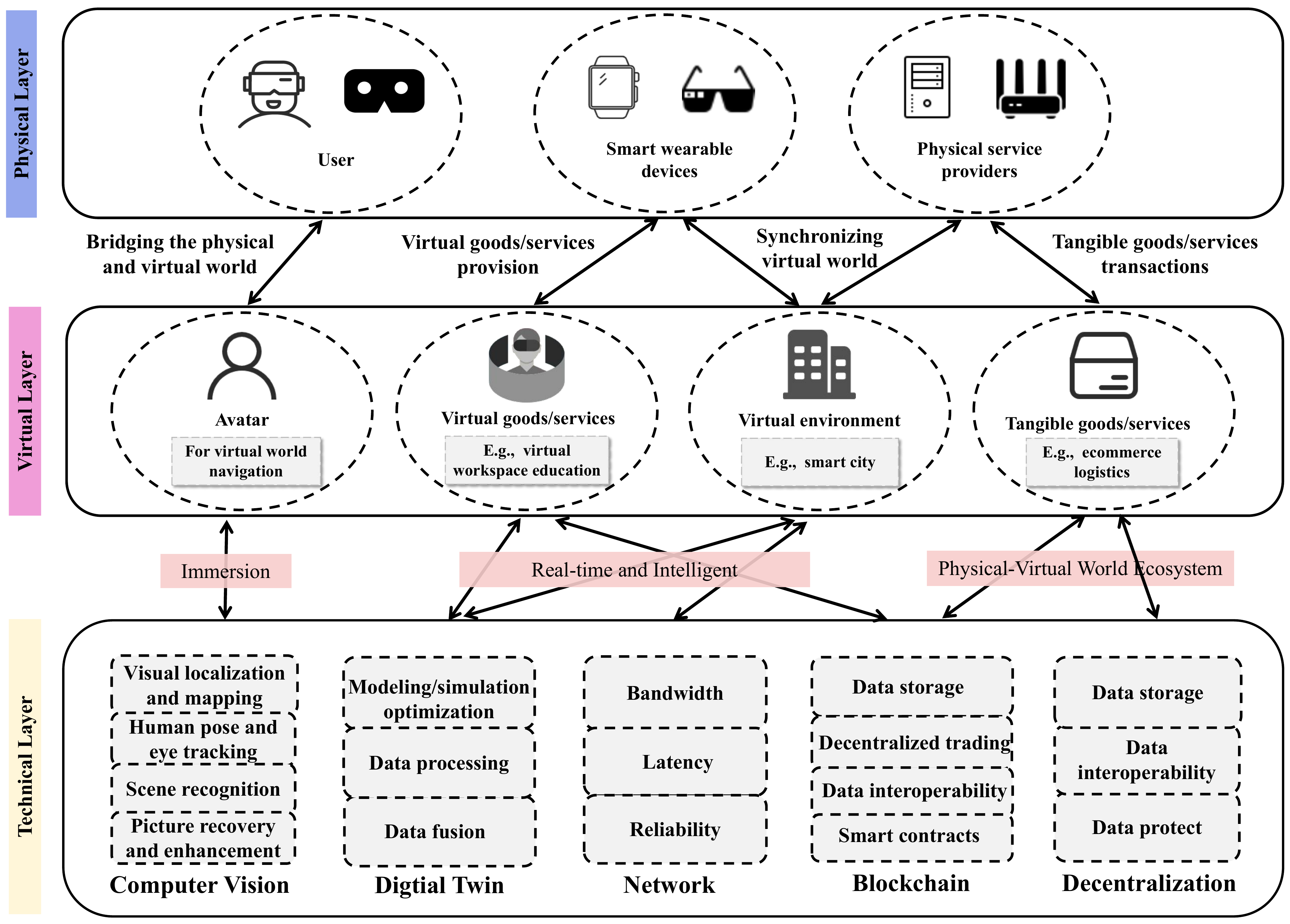}
	\caption{\textcolor{black}{The Metaverse architecture features the immersive and real-time physical-virtual world interaction supported by the Metaverse technical layer. The supporting technology ensures that the Metaverse is low bandwidth, low latency, enables ubiquitous access, and is trustworthy for users.}}\label{Metaverse}
\end{figure*}

\subsection{6G-enabled Edge Intelligence}\label{sec-2.2}
Edge artificial intelligence (Edge AI) \textsuperscript{\cite{8736011}} as a combination of AI and edge computing, which supports the deployment of machine learning algorithms to edge nodes to provide services to users, fully unlocking the potential of big data. 
Meanwhile, the continuous improvement of the computing power of edge nodes also provides a fundamental guarantee for the deployment of new 6G applications in the future\textsuperscript{\cite{2020Communication}}.
For example, in the 6G era of the Internet of Everything, edge intelligence has the potential to provide efficient and accurate intelligent navigation for vehicles by utilizing the powerful computing power of edge servers (e.g., roadside units)\textsuperscript{\cite{9199786}}.
Therefore, the advancement of edge intelligence allows the popularization and implementation of 6G technology in the real world.

As mentioned above, 6G edge intelligence has the advantages of low latency, computing offload, and high performance\textsuperscript{\cite{9502547}}. Therefore, the application of 6G-oriented edge intelligence has the following three benefits.

\textit{1) Balanced Data Storage:} Most of the data generated by terminal devices need to be sent to the cloud for processing in traditional cloud computing.
However, as the number of user devices increases, the workload of cloud servers becomes larger and larger.
Edge intelligence adopts load balancing technology (i.e., sending data generated by different terminal devices to edge storage nodes), reducing data storage redundancy.

\textit{2) Efficient Data Transmission:} The layered architecture of edge computing in the 6G ensures a shorter transmission time than cloud computing networks, significantly reducing the need for internet bandwidth\textsuperscript{\cite{8952884}}, increasing the speed of communication and information transmission, and providing a medium for the rapid response of virtual services\textsuperscript{\cite{lin2015leveraging}}.

\textit{3) High-Reliability:} In the future, 6G-oriented edge intelligence will use spatial multiplexing technology, where the probability of communication interruption is less than one in a million, which greatly improves the quality of user experience. 


Furthermore, from the perspective of model training, edge AI generally has the following three architectures:

\textit{1) Centralized Architecture:} Such an architecture means that the cloud server is responsible for the entire cycle of AI model training (including data collection, model training, and model inference). Specifically, remote terminal devices (such as smart watches, mobile phones, cars, and surveillance cameras) deployed at the edge layer generate and collect data required for AI model training. The cloud server then trains AI models on the data uploaded by these edge devices and deploys these models for model inference.

\textit{2) Decentralized Architecture:} Such an architecture means that each edge computing node is responsible for the entire cycle of AI model training. Specifically, each edge terminal device trains the AI model on its own local data. These edge computing nodes then exchange model information through network connections. Eventually these nodes get a shared global AI model by sharing local model updates. In this architecture, edge computing nodes can train AI models without the intervention of cloud servers.

\textit{3) Hybrid Architecture:} Such an architecture refers to a combination of centralized architecture and decentralized architecture patterns. Specifically, the edge server acts as the center of the architecture, and the edge terminal devices are responsible for local model training and sharing local model updates. More specifically, the edge server is responsible for optimizing the global AI model and using the network connection to distribute the updated AI model to edge nodes. At the same time, the training data of the edge nodes can also be uploaded to the cloud server for centralized training to give full play to the computing and resource advantages of the cloud server.

\section{METAVERSE}\label{sec-3}
\subsection{Metaverse Features}
{The Metaverse coexists and interacts with the real world, which is an online and shared virtual reality world.} In the Metaverse, users can get the same experience as the real physical world through customized 3D virtual avatars\textsuperscript{\cite{latoschik2017effect}}.
For example, avatars can perform a series of human activities in the Metaverse, such as shopping, telecommuting, and video conferencing. 
Specifically, the Metaverse has the following features:

\subsubsection{Immersive}
For the past decade, the way users interact with virtual worlds has been limited by screens and mobile devices. However, with the continuous development of Metaverse technology, the fusion between the physical world and the virtual world has largely alleviated this limitation. Specifically, in the Metaverse, the ``stimulus" received by the avatar can be fed back to the user in the physical world with high fidelity through XR technology and intelligent wearable devices such as brain-computer interfaces (BCI)\textsuperscript{\cite{schalk2004bci2000}}.
For example, users can wear virtual reality (VR) glasses and augmented reality (AR) headsets to enter the Metaverse for remote video conferences, realizing the transformation from the 2D application interface of the physical world to the 3D simulation space of the virtual world. It can improve the user's immersive experience\textsuperscript{\cite{segawa2020virtual,8643424}}.

\subsubsection{Multi-technology} 
The Metaverse contains a variety of advanced digital technologies and AI technologies, such as digital twin\textsuperscript{\cite{8972429}}, XR, blockchain\textsuperscript{\cite{feng2020joint}}, computer vision, and so on.
Specifically, digital twin technology enables large-scale, high-fidelity real-time mapping of physical entities, allowing users to simulate and test physical entities in the Metaverse\textsuperscript{\cite{el2018digital}}, XR technology uses mixed reality (MR) and AR to ensure the immersion of the Metaverse, and blockchain technology builds the economic system in the Metaverse.

\subsubsection{Interoperability}
Metaverse integrates multiple application scenarios, and some functions between the scenarios are interoperable\textsuperscript{\cite{wang2022survey}}.
For example, services such as virtual clothes and virtual feelings purchased by users can be used in various scenarios of the Metaverse, reflecting interoperability.

\subsubsection{Sociability}
The Metaverse is a new stage in the development of human social formations. 
Humans can pursue some higher-level needs beyond the physical world level in the Metaverse, such as virtual office, virtual entertainment, and other social behaviors. 
Therefore, sociality is an essential characteristic of communication between users in the Metaverse.

\subsubsection{Longevity}
As a digital world, the Metaverse is different from the physical world and has longevity characteristics. 
Specifically, the user's avatar, behavior, assets, and other data can be imprinted and stored in the Metaverse for a long time. 
However, in the real world, when an organization fails, or an individual dies, many things related to it will disappear with it. 
In contrast, users' personal information and digital assets are permanently stored in the Metaverse.


\subsection{Metaverse Architectures}
In this section, Fig. \ref{Metaverse} shows the architecture of the Metaverse, which includes physical layer, virtual layer, and technical layer. 
These architectures support the real-time interaction of users in the physical-virtual world.

\subsubsection{Physical Layer}
\textit{1) Users:} 
Users can interact with the virtual world of the Metaverse through intelligent devices such as VR, head-mounted displays (HMDs), AR goggles, and human gesture sensors\textsuperscript{\cite{xu2022full}}.
In the virtual world, users can immensely experience some applications, including custom identity, social interaction, entertainment, and learning\textsuperscript{\cite{ondrejka2004escaping}}. 
Specifically, in the Metaverse, users can redefine their digital image (i.e., 3D avatar), not limited by gender, appearance, and national boundaries. 
Meanwhile, users can immensely experience various virtualized scenarios in the Metaverse (e.g., virtual performances, virtual dating, 3D telecommuting) to meet their diverse needs.
Therefore, users interact in the physical-virtual world with the characteristics of diversification of information, identities, and modes.
These interactive behaviors ultimately give users an all-around immersive experience.

\textit{2) Smart Wearable Devices:} 
Users upload recorded data and historical behaviors from the real world to the Metaverse platform by smart wearable devices (e.g., smartwatches, smart wearable clothes, etc.). 
Meanwhile, the users can directly get some feedback from the avatar's behavior through the smart wearable devices, realizing the interaction between the physical world and the virtual world\textsuperscript{\cite{chen2017empirical}}.
Thus, smart wearable devices are an important medium for users to realize inter-operation between the physical and virtual worlds.

\textit{3) Physical Service Providers:} 
The physical service provider is the support and maintenance of the Metaverse by the technology of the physical world, including the communication at the edge of the network and the allocation of computing resources.
When users wear VR smart devices and enter the Metaverse, the physical service network operator will respond to the user's needs in the virtual world in real-time. For example, the reasonable allocation of network and communication resources is used to reduce the delay of interaction between the physical world and the virtual world\textsuperscript{\cite{lim2022realizing}}.
Moreover, the data generated by users in the Internet sensor network of the physical world will also be fed back to the virtual world for maintenance and utilization in real-time.
Therefore, the normal operation of the Metaverse cannot be separated from the support of physical service providers.



\begin{table*}[!t]
\scriptsize
	\centering
	\caption{\textcolor{black}{Technologies that support the Metaverse construction.}}
	\begin{tabular}{|c|c|}\hline
		\textbf{Technology}&\textbf{Support for the Metaverse}\\\hline
		Digital twins&Copy entities from the physical world into the virtual world of the Metaverse.\\\hline
        Blockchain&Solve the problems of limited storage resources and privacy leakage in a centralized architecture.\\\hline
        Communications network&Make the network have the characteristics of high bandwidth, low latency, and reliability.\\\hline
        Decentralization&Ensure user data security and system robustness.\\\hline
        Computer vision&Guarantee the XR technology necessary for the Metaverse.\\\hline
	\end{tabular}
	\label{tab-1}
\end{table*}


\subsubsection{Virtual Layer}
As shown in Fig. \ref{Metaverse}, the virtual layer of the Metaverse includes virtual environments, virtual services, virtual currencies, and all behaviors of avatars in the virtual world.
In the virtual layer, technologies such as high-definition rendering and simulation are used to develop and maintain the virtual spaces of the Metaverse (such as social networks, economic systems, and business activities) so that users can have an immersive, high-quality experience\textsuperscript{\cite{9097455}}.
Similar to shopping in a real-world mall, users can purchase virtual goods such as paid videos and clothing from good providers in the Metaverse for an interactive experience\textsuperscript{\cite{liu2019decentralized}}.
\subsubsection{Technical Layer}
In this section, we introduce some techniques applied to the Metaverse. A summary has been given in Tab. \ref{tab-1}.
Next, as shown in Fig. \ref{Metaverse}, we describe the support each technology provides to the Metaverse.
\begin{itemize}
    \item \textit{Digital Twin:} The digital twin is a technique that can convert physical entities to virtual and digital models through information technology. In the process of modeling, digital twin technology can map the composition, characteristics, functions, and performance of physical entities into the Metaverse in real-time\textsuperscript{\cite{wu2021digital}}. 
    Thus, digital twins and physical entities in the Metaverse are nearly equivalent\textsuperscript{\cite{liu2019novel}}. 
    This largely facilitates the fusion of the virtual and physical worlds.

    \item \textit{Blockchain:} All objects in the Metaverse, such as digital twins, avatars, and complex maps, exist as digits. 
    However, this inevitably requires servers' storage capacity because of massive data. 
    Meanwhile, the traditional centralized architecture cannot upload all the data to the cloud servers due to limited network resources.
    On the other hand, the Metaverse system based on centralized architecture may lead to some risks such as monopoly and illegal control\textsuperscript{\cite{2018A}}. 
    To be more specific, a company in a monopoly position holds a large amount of personal data about its users. 
    Once the company's servers are attacked, many users' privacy will likely be leaked.
    Blockchain, where data are stored in blocks in chronological order and the blocks are sequentially linked to form a chained data structure, is a distributed shared database\textsuperscript{\cite{8489897,8624307}}.
    More importantly, blockchain also has its characteristics of being tamper-proof and non-modifiable\textsuperscript{\cite{mozumder2022overview}}. 
    Thus, the distributed database stored in the blockchain can effectively alleviate the problems of limited storage resources and privacy leakage, which the centralized architecture faces in the Metaverse.
    
    \item \textit{Network Communication:} In the Metaverse, users' immersive experience, virtual office, and perceptual communication rely on high bandwidth, low latency, and high-reliability network communication technologies.
    Specifically, VR devices require a transmission rate beyond 250 $Mbit/s$ and data error rate between $10^{-1}$ and $10^{-3}$ to prevent communication interruptions\textsuperscript{\cite{lincoln2016motion}}. 
    Haptic communication requires a lower data transfer rate at least $1$ $Mbit/s$ and a latency of no more than 1 $ms$\textsuperscript{\cite{ivkovic2015quantifying}}. 
    These requirements can be achieved with enhanced mobile broadband and ultra-reliable and low-latency communication technologies\textsuperscript{\cite{rappaport2019wireless}}.
    
    \item \textit{Decentralization:} 
    In the Metaverse, due to security and privacy issues, it is not allowed to store all users' data on a single server. 
    It easily causes the whole Metaverse system to go down, and the user's private data leaked because of the server's breakdown. 
    In contrast, in the decentralized storage system, every involved organization can become a storage node\textsuperscript{\cite{frey2008solipsis}}. 
    Therefore, the absolute power ``king'' of the Metaverse does not exist in the decentralized storage system, and no one can have absolute power to control all users' data. 
    Even Meta, a giant technology company, can only act as a builder in the Metaverse.
    In particular, many companies are all representatives of decentralized storage systems in the Metaverse, such as Filecoin, Storj, Arweave, and MEM. 
    Specifically, they all dispersedly store slices of data on multiple independent network nodes and then establish an economic incentive mechanism to enable users to store data for a long time\textsuperscript{\cite{xu2021wireless}}.
    
    \item \textit{Computer Vision:} 
    Computer vision is a technology that interacts between the digital world and the real world, laying the foundation for image processing for the realization of XR\textsuperscript{\cite{huynh2022artificial}}.
    Some applications in the Metaverse, such as visual localization and mapping, human pose and eye tracking, overall scene recognition and understanding, and picture recovery and enhancement, mainly apply computer vision. 
    For instance, in Fig. \ref{Metaverse}, visual localization and mapping are like human beings in the physical world perceive everything with their eyes and then map them to the brain to reconstruct a three-dimensional world.
    Similarly, in the Metaverse, the avatar also needs to obtain the three-dimensional structure of the unknown environment and perceive the exact movement position of the object.
\end{itemize}


\subsection{Metaverse Applications}
There are some application scenarios for Metaverse, such as entertainment and game industry, immersive education, smart city, medical field, industrial simulation testing, digital tourism, etc. In this section, as shown in Fig. \ref{application}, we present five common application scenarios in the Metaverse.

\subsubsection{Immersion Education}
Metaverse builds immersive education scenarios through digital twin technology. 
Specifically, students can promote their understanding of learning content through immersive learning in virtual scenarios.
Immersive education already has many application platforms in the Metaverse. 
For example, Hoodoo Labs\textsuperscript{\cite{hong2017english}} is a learning English platform that has transposed more than 300 characters and above 4,300 scenarios that English conversations can apply to virtual reality scenarios.
And then, users can improve their English skills through immersion, even though they travel to any continent or village. 
Moreover, Virbela, as a virtual world platform created to address collaboration in distance learning courses, was used by Davenport University to create a custom virtual campus named ``Davenport Global" during the COVID-19 pandemic.
Specifically, ``Davenport Global" preserves the physical campus's classroom culture and practical experiences, such as interactive auditoriums, presentation screens, private tutorial rooms, and free space for socializing and learning, enabling students to gain a sense of belonging just like a real campus.

\subsubsection{Telecommuting}
\begin{figure}[!t]
	\centering
	\includegraphics[width=0.75\linewidth]{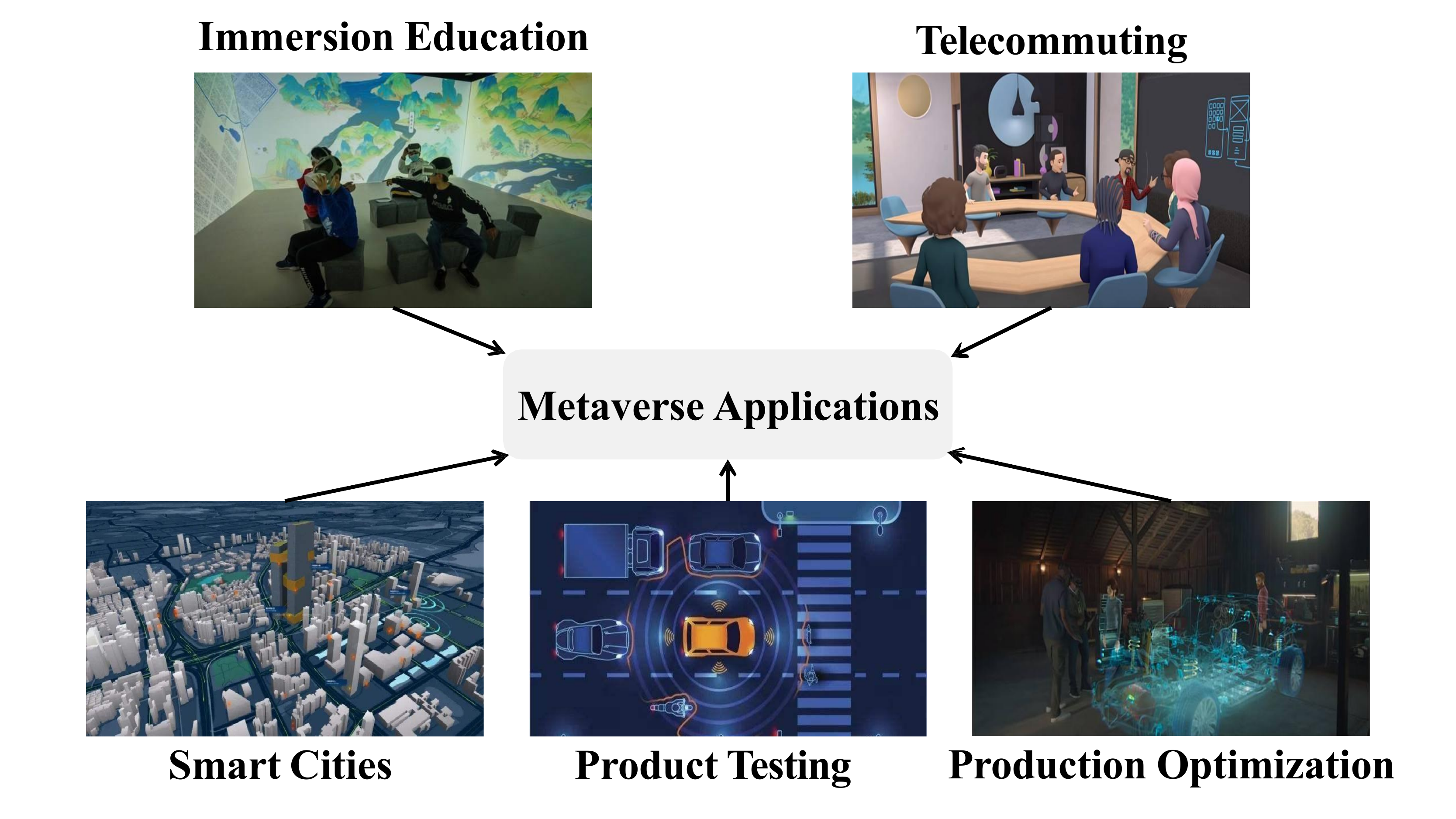}
	\caption{\textcolor{black}{Common application scenarios in the Metaverse.}}\label{application}
\end{figure}

Under the influence of COVID-19, many applications, such as telecommuting and videoconferencing, have gained universal popularity in the latest two years. 
However, business-to-business (B2B) or customer-to-customer (C2C) videoconferencing brings some problems such as monotony and lack of spatiality. 
The emergence of the Metaverse has enriched the telecommuting and videoconferencing scenarios. 
Specifically, in Fig. \ref{application}, the offices, conference rooms, and tables have been built in the Metaverse to solve the lack of spatial sense. 
The monotony problem has also been solved by using spatial audio technology to simulate footsteps, live sounds, etc.

\subsubsection{Smart Cities}
The Metaverse applies digital twin technology to mapping the real-world cities into the virtual world in real-time\textsuperscript{\cite{ruohomaki2018smart}}. 
In detail, the digital twin technology can digitize a series of physical objects such as traffic roads, city buildings, and vehicles in real-world cities to construct virtual cities. 
Conveniently, researchers can conduct simulation experiments in the virtual world to achieve facilities and optimize the allocation of resources in real cities.
For example, the intelligent parking solution based on the Metaverse platform built by Intel and Siemens in Berlin has brought great convenience to local traffic management\textsuperscript{\cite{yan2011smartparking}}.

\subsubsection{Product Testing}
Before a product can be put on the market, it usually needs to be fully and complexly tested in a simulation environment.
As a parallel, virtualized, and digital open world, the Metaverse provides natural advantages for product testing. 
In detail, the product can be tested simultaneously in physical space and virtual space through the combination of virtual and real, so as to reflect more intuitively the changes of the product to improve the efficiency of product testing and certification\textsuperscript{\cite{DBLP:journals/corr/abs-2201-01634}}. 
For example, the industrial Metaverse provides a virtual test space for automotive-grade chips, which allows engineers to enter the Metaverse to simulate and test the chips (e.g., simulating and testing the safety of self-driving cars equipped with AI chips). 
In this way, it not only saves costs but also improves the testing efficiency of automotive-grade AI chips.

\subsubsection{Production Optimization}
Metaverse can optimize the industrial structure of factories to improve production efficiency\textsuperscript{\cite{han2022dynamic}}.
Specifically, in the early stage of smart factory construction, digital twin technology can map the factory building structure, production line layout, and production process to the Metaverse in real-time.
Afterward, workers can adjust the factory's capacity, equipment, and staffing for maximum production efficiency in the Metaverse. 
For example, BMW's introduction of Nvidia's Metaverse platform Omniverse to coordinate the production of cars at 31 factories is expected to improve production planning efficiency by nearly 30\%\textsuperscript{\cite{alkazzi2020leveraging}}.

\begin{figure}[!t]
	\centering
	\includegraphics[width=1\linewidth]{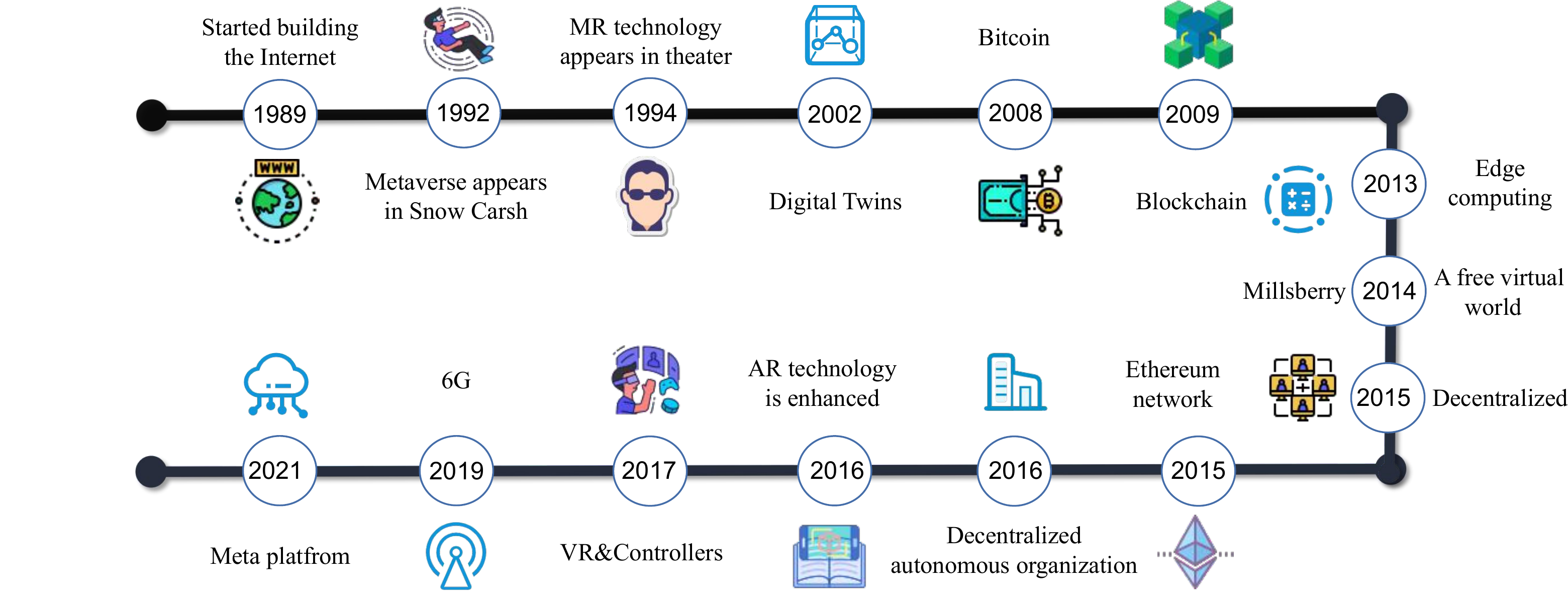}
	\caption{\textcolor{black}{The development in the Metaverse from the perspective of technological updates.}}\label{development}
\end{figure}

\subsection{Metaverse History}
Fig. \ref{development} shows the development timeline of the Metaverse.
The term ``Metaverse" was coined in Neal Stephenson's science fiction novel Snow Crash in 1992. The novel describes a virtual world detached from the real world, but parallels with the real world interact with each other and are always online.
Since then, in 2003, the popularization of the Internet and the development of digital twin technology made the construction of the Metaverse platform possible.
At this time, the American Linden Laboratory developed the ``Second Life" online game. 
In this game, players can make whatever they want in their field. However, due to the lack of immersion in Second Life, it is only an initial foray into the realm of the Metaverse. 
With the birth of Bitcoin in 2008 and blockchain technology in 2009, the Metaverse established an independent economic system.
In 2014, the development of VR technology had realized the transformation of the Metaverse from ``planar, passive and one-way" to ``three-dimensional, active and interactive." 
Especially since 2016, with the popularization of VR terminal devices represented by Oculus and HTC Vive, the sensory stimuli (such as vision, hearing, touch) generated by the user's avatar in Metaverse can be transformed into reality through VR equipment and somatosensory.  These VR devices enhance the user's sensory experience, thereby greatly enhancing the user's sense of immersion.
In the same year, the maturity of decentralization technology also caused some companies to shift from developing centralized architecture Metaverse platforms to decentralized Metaverses.
In 2019, the continuous development of network communication technology and AI had reduced the communication overhead of VR devices and the delay of user interaction in the real-virtual world, greatly enhancing the user's immersive experience.
In 2021, Roblox wrote the Metaverse concept into its prospectus and successfully landed on the New York Stock Exchange. Meanwhile, Facebook founder Mark Zuckerberg announced that the company would be renamed Meta. 
Since then, the development of the Metaverse has ushered in another wave.

\section{6G-ENABLED EDGE AI EMPOWERED METAVERSE}\label{sec-4}
\subsection{Edge Intelligence-Based Metaverse Architectures}

\begin{figure}[!t]
	\centering
	\includegraphics[width=0.75\linewidth]{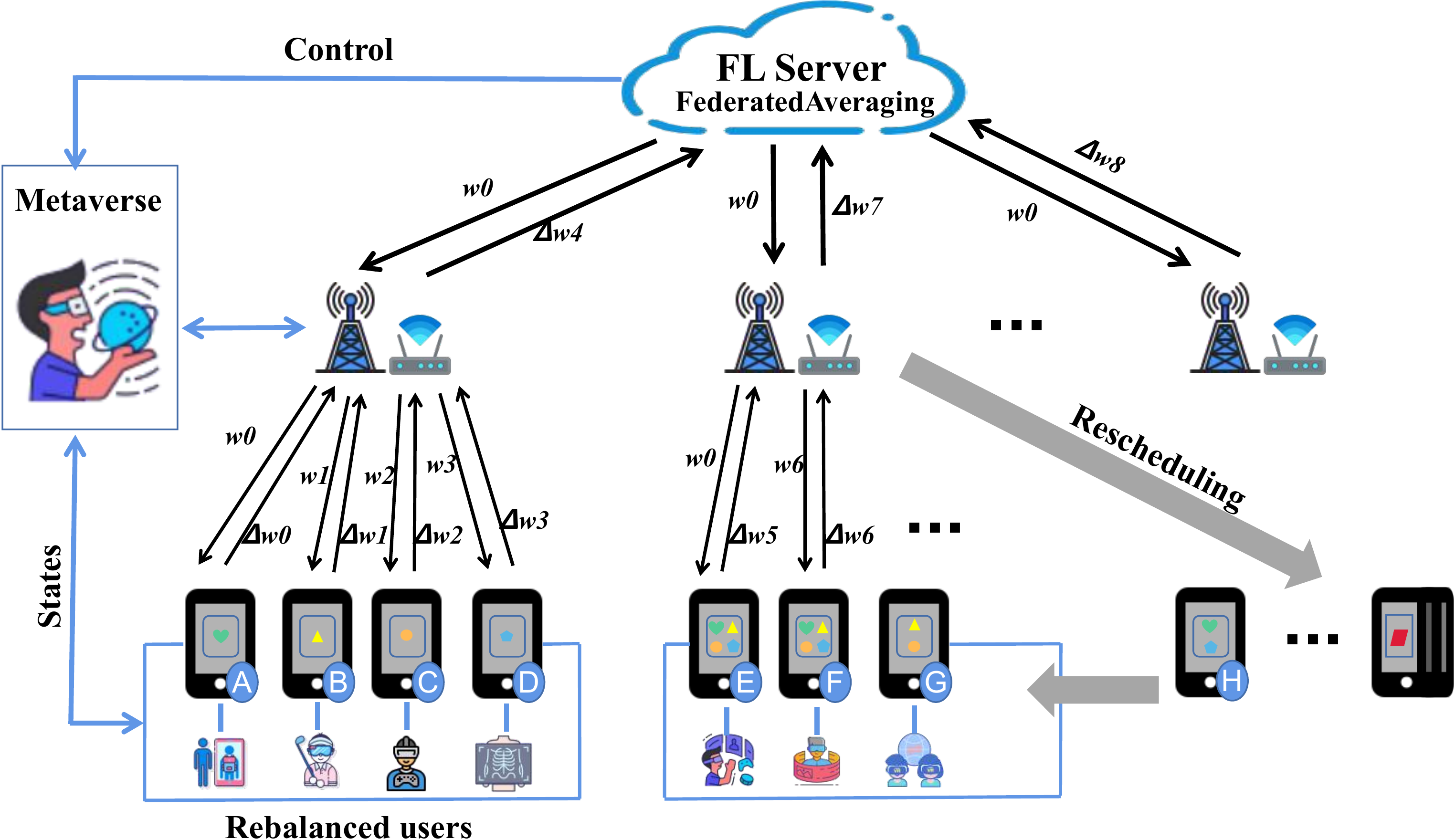}
	\caption{\textcolor{black}{The Metaverse architecture of self-balancing federated learning.}}\label{edge cloud}
\end{figure}


\subsubsection{Edge Cloud-Metaverse Architecture}
In the Metaverse, the traditional server-centric network architecture needs to transmit large amounts of data between cloud and terminal devices. 
However, due to the large-scale multiplexing nature of the network, the available bandwidth used to transfer data between the cloud server and terminal devices can change dramatically over time. 
This leads to many problems such as video frame drops and high latency, which degrade the quality of user experience\textsuperscript{\cite{zhang2017towards}}. 
As a distributed, cloud-centric, and edge-cooperative integrated network, the cloud-edge network integrates terminal devices, edge computing nodes, and cloud servers with high real-time performance, efficient communication, and stable security. 
Therefore, in reference \cite{younis2020latency}, the cloud-edge network architecture was introduced into the Metaverse to solve the problems of insufficient bandwidth and high latency.

Although the edge cloud-Metaverse architecture has advantages of high real-time performance, efficient communication, and stable security, it also has limitations. For example, users in different scenarios of the Metaverse may generate data with different features and AI tasks with diverse needs, resulting in an imbalance in the distribution of computing power of edge nodes.
To this end, we propose a self-balancing federated learning-based Metaverse framework to address the statistical heterogeneity faced by edge-cloud architectures.
Federated learning (FL) consists of a server and multiple clients, in which the server delivers the global model and aggregates the client's local model, and the client trains the received global model\textsuperscript{\cite{liu2020privacy,9663107,liu2020secure,8994206,8832210}}.
In the proposed architecture, edge nodes also play the role of intermediaries in addition to the tasks of computational offloading and storage cache. As shown in Fig. \ref{edge cloud}, when these users (i.e., clients) collaborate together to train a high-quality global model, the edge node will plan the edge nodes that the user needs to access to achieve local balance according to the data distribution or AI task type of each user\textsuperscript{\cite{9141436}}.
Next, the model trained by each user using the local dataset is uploaded to the FL server through edge nodes for average aggregation.
The entire FL training process is repeated iteratively until the global model converges.

\begin{figure}[!t]
	\centering
	\includegraphics[width=0.75\linewidth]{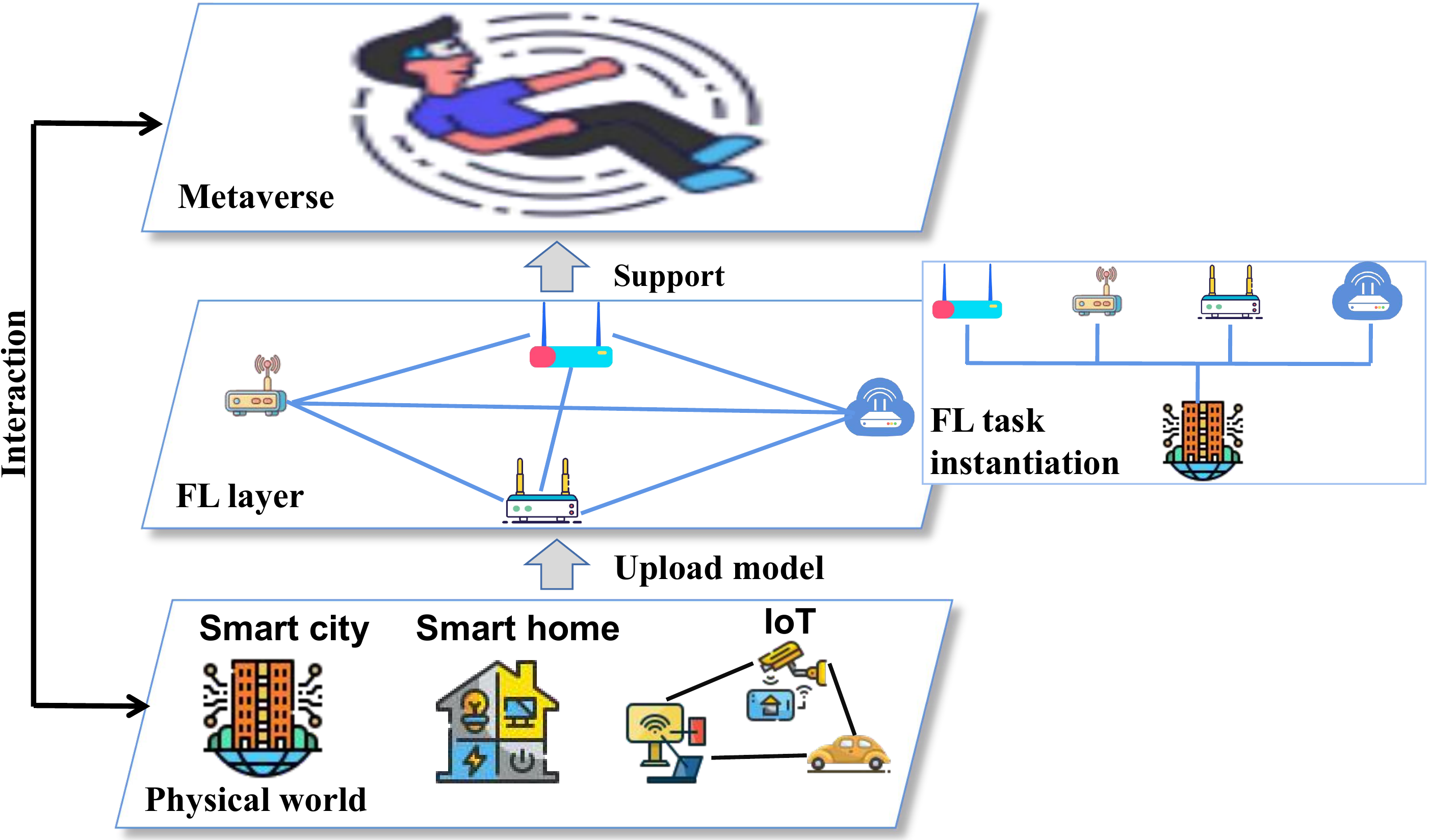}
	\caption{\textcolor{black}{Overview of the self-balancing federated learning-based Metaverse framework.}}\label{mobile edge cloud}
\end{figure}

\subsubsection{Mobile Edge Cloud-Metaverse Architecture}
Mobile edge computing (MEC), as an emerging paradigm for edge computing, has differed from traditional cloud computing in that cloud services can run as close to users as possible to reduce network latency\textsuperscript{\cite{8424113}}.
MEC is applied in some highly mobile devices, such as the Internet of Vehicles, mobile terminals, wearable devices, etc., to reduce network delay has become a new hot direction\textsuperscript{\cite{corneo2021surrounded}}.
For instance, Zhang et al.\textsuperscript{\cite{zhang2017towards}} introduced the MEC into the Metaverse to improve the quality of users' experience. 
Specifically, in the Metaverse of the MEC architecture, dynamic edge nodes can be as close as possible to users to combine multiple edge nodes to assist in completing the same user's instructions, effectively solving the delay problem caused by user mobility.

However, when multiple edge nodes assist in cooperation, it is necessary to transmit the user's data, which may lead to leakage of user privacy and serious consequences such as identity crisis in the Metaverse.
Therefore, in this paper, we propose an FL-based Metaverse of MEC architecture for protecting users' privacy.
Specifically, model rather than user data is transmitted between edge nodes in this architecture.
In addition, in Fig. \ref{mobile edge cloud}, the edge nodes located at the FL layer can also construct a directed acyclic graph (DAG) to improve the computational efficiency of the system\textsuperscript{\cite{9653818}}.

\subsubsection{Decentralized Metaverse Architecture}
The Metaverse allows users to access virtual worlds through mobile smart devices, enabling real-time gameplay anywhere. 
However, the Metaverse development brings scalability issues to the basic physical facilities (e.g., cloud servers, edge computing nodes). 
In other words, as the number of online users of the Metaverse increases, the infrastructure needs to take on more computing tasks. 
To maintain a good user experience quality, the server must quickly process and respond to requests from more terminals in a short period. However, this is generally impractical for the Metaverse systems under traditional centralized architectures. 
Therefore, references \cite{frey2008solipsis,liu2019decentralized,xu2021wireless,nguyen2021metachain} proposed some blockchain-based Metaverse systems to solve the above problems.
Due to the independence and security of the blockchain, reference \cite{nguyen2021metachain} describes how to use blockchain to solve the scalability problem in the Metaverse.
Reference \cite{xu2021wireless} designed an incentive mechanism to improve the transaction efficiency in the Metaverse by combining the Dutch double auction mechanism and reinforcement learning.

\begin{figure}[!t]
	\centering
	\includegraphics[width=0.75\linewidth]{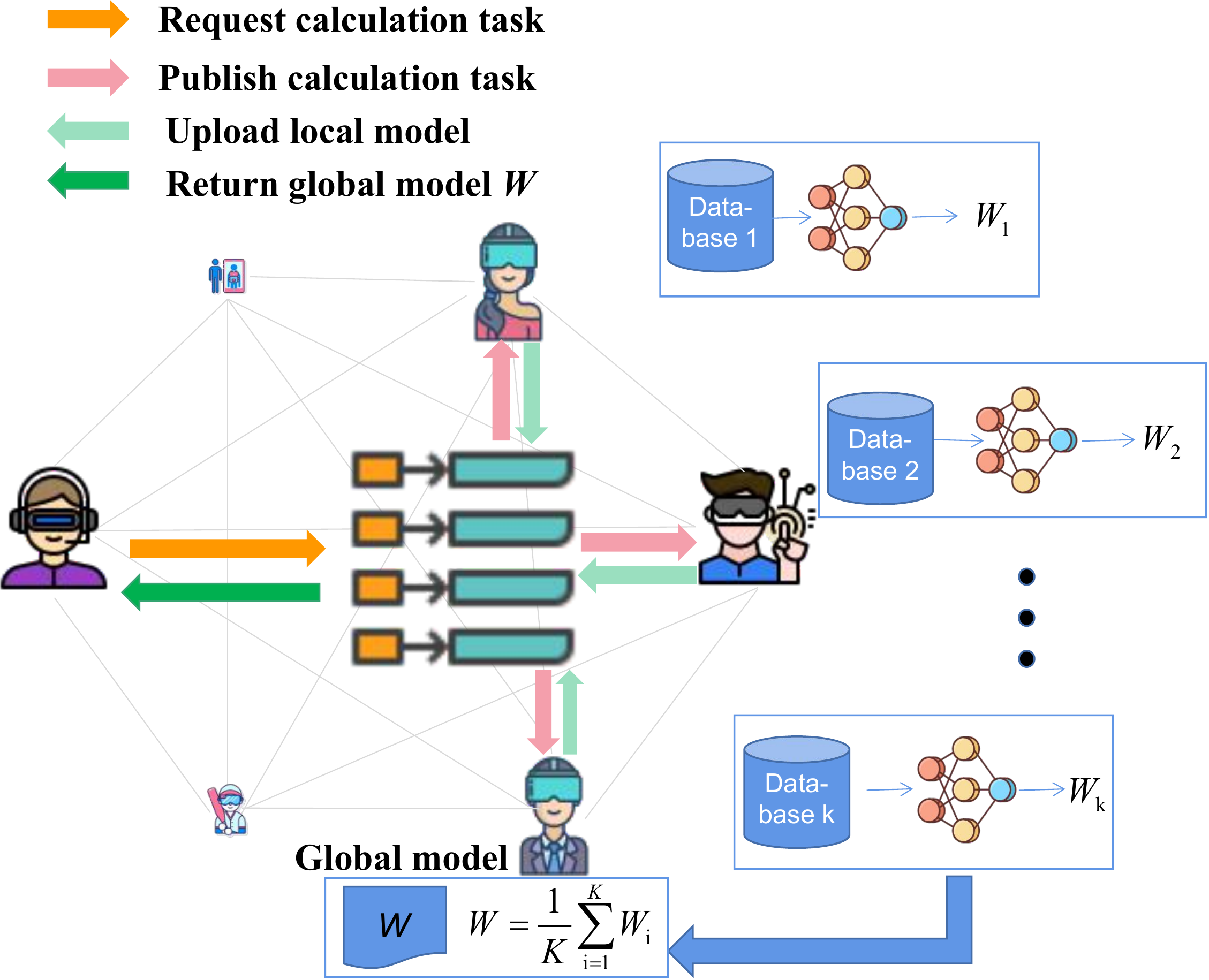}
	\caption{\textcolor{black}{A decentralized Metaverse architecture based on federated learning combined with blockchain.}}\label{blockchain}
\end{figure}

However, the above methods suffer from the insufficient performance of the trained models. Specifically, in a blockchain-based decentralized Metaverse system, users can only interact with each other to train models without effectively utilizing as much user data as possible due to strict privacy restrictions.
We propose a decentralized Metaverse architecture based on federated learning combined with blockchain for this problem.
As shown in Fig. \ref{blockchain}, a user can request a computing task through the blockchain, and then users with similar data distribution will receive the task request broadcast by the blockchain.
Next, these users upload their local models to the blockchain, and the blockchain aggregates the received models on average and returns them to the task requester.
Finally, the entire transaction process will be recorded on the blockchain for traceability.
Therefore, under this architecture, the local model is optimized through the cooperation of multiple user nodes, which effectively solves the problem of insufficient model performance in the blockchain-based decentralized Metaverse.

\subsection{Challenges and Advanced Methods}
Although the edge intelligence with 6G has been applied to solve problems such as high bandwidth and high connection density in the Metaverse. 
However, the Metaverse still faces many challenges, such as users' privacy, network latency, and resource allocation issues. 
In this section, we introduce important advanced methods to address these challenges (as shown in Table. \ref{tab-3}).

\begin{table*}[!t]
\scriptsize
	\centering
	\caption{\textcolor{black}{Summary of the challenges and corresponding advanced methods.}}
	\begin{tabular}{|c|c|c|}\hline
		\textbf{Reference} & \textbf{Issue} & \textbf{Main Idea}\\\hline
		
		\cite{falchuk2018social} &Privacy & Creating multiple user replacement avatars to obfuscate the user's real location and real target. \\\hline
		
		\cite{9170905} &Privacy & Introducing federated learning to protect local user data. \\\hline
		
		\cite{xu2021wireless} & Delay & Proposing an auction mechanism based on deep reinforcement learning  to improve communication efficiency. \\\hline
		
		 \cite{jiang2021reliable} & Delay & Designing a blockchain-based incentive mechanism to motivate workers to actively perform computing tasks. \\\hline
		 
		 
		 \cite{nguyen2021metachain} & Resource Allocation & Leveraging shard technology to create shards based on actual user needs. \\\hline
		 
		 \cite{9687500} & Resource Allocation & Recording of the resource requirements of different users to select the appropriate model training solution by edge nodes. \\\hline
	\end{tabular}
	\label{tab-3}
\end{table*}

\subsubsection{Privacy}
With the continuous development of various communication and virtual technologies, the Metaverse provides users with excellent experiences in some virtual scenarios such as entertainment games and smart cities.
However, the Metaverse is a virtual world that is interconnected with the real world and exists in parallel, and it has the same digital information security issues as the real world.
In reference \cite{leenes2007privacy}, the authors proposed that there is a privacy problem in the Metaverse.
Metaverse can potentially monitor users' physiological responses and body movements, thereby leaking sensitive personal information such as user habits and their physiological characteristics to third parties.
In addition, in the virtual world, evil users can also imitate other users to obtain other people's personal information.
In large-scale online games, some matters such as user harassment and theft of other people's finances threaten users' privacy and security all the time\textsuperscript{\cite{falchuk2018social}}. 

Considering the risk of private data leakage of users, the authors in reference \cite{falchuk2018social} proposed two privacy protection mechanisms ``clone cloud" and ``private copy". 
The ``clone cloud" creates multiple users' substitutes in the system and makes the substitutes move continuously in the Metaverse environment to confuse the users' location information, activities, goals, and other information. 
This approach can make the Metaverse environment into chaos so that intruders cannot distinguish the real users. The ``private copy" allows users to apply for virtual private storage space in the Metaverse, and the entire experience process takes place in the private space rather than in the complete virtual world that is widely accessed, effectively protecting the user's privacy.
Besides, the authors in reference \cite{9170905} proposed a blockchain-based digital twin wireless network (DTWN) edge computing federated learning framework to solve the problem of user privacy data security.
The blockchain was introduced to improve the security of the untrusted user's digital twin because it can ensure the security of the digital twin data, and the user's part model can be stored in it. 

\subsubsection{Delay}
One of the characteristics of the Metaverse is an immersive virtual experience for users. 
However, a high-speed and low-latency network connection is an important foundation for improving the user experience. 
In the virtual world created by the basic digital technology, the user may experience visual jitter or delay and other undesirable phenomena, which seriously affect the user's immersive experience due to insufficient network bandwidth.

To reduce network latency, an incentive mechanism framework for VR services was proposed in reference \cite{xu2021wireless}, which uses perceived quality as a criterion for measuring immersive experience and effectively evaluates the immersive experience in the Metaverse.
Meanwhile, the authors designed a double dutch auction mechanism based on deep reinforcement learning (DRL), which not only improves the communication efficiency and reduces the auction cost but also realizes the dynamic pricing of users' immersive services in the virtual world, ensuring the rationality of the incentive mechanism and the authenticity. 
Jiang et al.\textsuperscript{\cite{jiang2021reliable}} found that coded distributed computing (CDC) can improve the latency problem in the Metaverse and proposed a CDC and dual blockchain distributed collaborative computing framework. 
The CDC is responsible for aggregating the idle resources of devices connected to the Metaverse to process the more intensive computing tasks in the Metaverse.
The dual-blockchain measures CDC’s worker reliability by estimating future rewards through the workers' historical records. 
Furthermore, the authors design an incentive mechanism (i.e., optimal computing rewards can be obtained for devices assigned to large-scale computing tasks) to motivate CDC workers to actively participate in the computing tasks of the Metaverse to improve the computing speed. 


\subsubsection{Resource Allocation}
The key to the connection between the real world and the virtual world is collecting data in the physical world and updating it in real-time in the virtual world so that users can have a real and lasting interactive experience.
However, the computing, communication, and storage shortage will seriously affect the user's immersive experience. 
At present, in the Metaverse, reasonably allocating resources and effectively coordinating work among computing nodes are still essential challenges. 

For the resource allocation problem, a new blockchain-based framework called Metachain was proposed in reference \cite{nguyen2021metachain}. 
Metachain introduces a new sharding technology, which can create shards according to users' actual computing resource requirements, which was used so that service providers can dynamically allocate computing resources to shards. 
Moreover, to meet the demand for computing resources in the Metaverse, the author uses Stackelberg game theory\textsuperscript{\cite{1934Marktform}} analysis to propose an incentive mechanism, i.e., users obtain corresponding rewards by providing resources to blockchain shards.
Based on the intelligent 6G edge network, a machine learning framework was proposed in reference \cite{9687500} for decentralized learning and coordination of edge nodes to improve resource allocation strategies. 
Specifically, during the training process, the nodes select an appropriate model training scheme by recording the different resource requirements of different users.
One of the solutions is the computational offloading and resource allocation algorithm of multi-agent DRL, i.e., the local model is cooperatively trained using local information sharing.
Another solution is the resource allocation algorithm of the federated framework based on DRL, i.e., the edge node uploads the model parameters to the server to update the global model after completing a round of training.

\section{OPEN RESEARCH TOPICS AND FUTURE DIRECTIONS}\label{sec-5}
\subsection{Perceptual Realization}
One of the essential goals of the Metaverse is to provide users with a multi-sensory immersive experience. 
The user's immersive experience mainly depends on the effectiveness of wearable devices such as VR and mobile headsets. 
However, wearable devices only utilize the user's visual, auditory and tactile perceptions when interacting with the virtual world. 
Besides, feedback from the user's brain is also essential. 
BCI is a communication channel between brains and external devices, which can realize the exchange of information between the brain and the device and then operate the devices through the ideas generated by the brain, which can significantly improve the user's immersive experience quality.
Specifically, BCI technology can be divided into invasive and non-invasive. 
The invasive BCI is implanting deep electrodes into the brain to obtain accurate signals by implanting signal acquisition equipment. 
For the non-invasive BCI, although it is convenient to wear, there is a specific error in recognizing brain waves. 
Therefore, striking a balance between precision and convenience will be an essential research direction for applying BCI technology in the Metaverse.

\subsection{Code of Ethics}
Ethical issues in the Metaverse arise from conflicts between the behavior of avatars in the virtual world and the moral norms of the real world.
Specifically, the Metaverse redefines ``second identities'' (i.e., avatars) for people and provides them with free virtual space. 
However, as more and more users immerse themselves in the Metaverse and enjoy virtual lives, the unfettered behavior of avatars brings about more complex social relationships than in the real world. 
For example, car grabs and crashes in the GTA5 game are usually condemned in the real world and may be permissible in the virtual world. 
Therefore, such behaviors that are contrary to real-world moral norms may adversely affect minors who are not yet mature\textsuperscript{\cite{slater2020ethics}}. 
Thus, as a next-generation network, the Metaverse must control and constrain the behavior of users and establish explicit ethical and moral norms to maintain a good and orderly ecological environment of the Metaverse\textsuperscript{\cite{ning2021survey}}.

\subsection{Security}
One of the main characteristics of the Metaverse is providing users with an immersive experience. 
The immersive experience mainly relies on the intelligence of wearable devices such as VR and mobile headsets.
In the user's daily use, wearable devices can read and store the user's fingerprints, iris, habits, voice, and other personal identity data. 
Then, those data will be uploaded to the Metaverse servers to meet the user's immersive experience needs. 
However, when the server of the Metaverse platform fails or is attacked, it may lead to the users' identity information leakage, which may cause some identity crisis issues, such as scandals caused by false avatars, identity theft, and body swapping. 
Therefore, it is necessary to strengthen the security of the Metaverse system and improve the ability to protect users' identity data. 
For security, due to the blockchain adopting a consensus mechanism and the encryption algorithm, it ensures data security, immutability, and transparency. 
Therefore, the Metaverse system based on blockchain authorization will be an effective solution. 

\subsection{Balance Between Virtual and Reality}
Metaverse is a virtual space based on the real world, which builds platforms for social interaction, life, and entertainment in the virtual world, and realizes the integration of the real world and the virtual world.
Just as in the movie Ready Player One, users can enter the virtual space through VR and wearable devices to get an immersive experience. Specifically, through the somatosensory clothing, the user can feel the pain of being attacked by the body; through the fully automatic haptic chair, the user can experience the somatosensory feeling of falling and flying in the virtual world. 
However, users who are excessively indulged in the Metaverse may not be able to distinguish between the virtual and the real, which will seriously affect their work, life, and study in reality.
Thus, it is a future research topic to achieve a balance between virtual and reality so that users can obtain a high-quality virtual experience without indulging in it and affecting normal life. 
To this end, we propose three schemes for researchers' reference: 1) Limit users' access age and usage time. 2) Reduce the realism of the virtual world to prevent users from becoming addicted. 3) Set up a regulatory body to scrutinize whether virtual functions will have undue negative effects.

\section{CONCLUSION}\label{sec-6}
In this survey, we introduced the integration of 6G-oriented edge intelligence into the Metaverse. We investigated three Metaverse systems based on 6G edge intelligence.
Furthermore, we discuss some of the core challenges in the Metaverse.
We provided a comprehensive introduction of the advanced methods for the above challenges.
Finally, we present some open topics and directions for future research in the Metaverse.
The survey serves as the initial step that precedes a comprehensive investigation of the Metaverse based on edge intelligence with 6G and hopes to offer insights and guidance for the researchers and practitioners to expand on the Metaverse.





\bibliography{ref}




\end{document}